\documentclass[preprint, draft, prb, endfloats,API,pre,doublespace]{revtex4}
\usepackage{amsmath}
\usepackage{amssymb}
\usepackage{graphicx}
\usepackage{dcolumn}
\usepackage{bm}

\begin{document}
\title{spin polarization direction switch based on an asymmetrical quantum wire}
\author{Zhonghui Xu$^{1,2}$, Xianbo Xiao$^{3}$ and Yuguang Chen$^{1,a)}$\footnote[0]{$^{a)}$corresponding author: ygchen@tongji.edu.cn}}

\address{\\$^{1}$Department of Physics, Tongji University, Shanghai 200092, China.\\
   $^{2}$Faculty of Information Engineering, Jiangxi University of Science and Technology, Ganzhou 341000, China.\\
  $^{3}$School of Computer, Jiangxi University of Traditional Chinese Medicine, Nanchang 330004, China.}

\begin{abstract}

A scheme for a spin polarization direction switch is investigated by studying
the spin-dependent electron transport of an asymmetrical quantum wire (QW) with
Rashba spin-orbit coupling (SOC). It is found that the magnitude of the
spin-polarized current in the backward biased case is equal to
it in the forward biased case but their signs are contrary. This
results indicate that the spin polarization direction
can be switched by changing the direction of the external current. The
physical mechanism of this device may arise from the symmetries in the
longitudinal and transverse directions are broken but
$C_2$-rotation and time-reversal symmetries are still reserved. Further
studies show that the spin polarization is robust against disorder, displaying
the feasibility of the proposed structure for a potential application.

\end{abstract}

\maketitle

\newpage

In the past decades, spin-dependent electronic transport in low-dimensional
mesoscopic systems has been investigated extensively both for fundamental
physics and for potential applications in spintronic devices$^{1}$ , in
which the electron spin degree of freedom may be used as well as its charge
for information processing. The basis of this application is the generation
of spin-polarized current and quantum control of coherent spin states. The
Rashba SOC$^{2}$ plays an important role in
spin-dependent electronic transport. SOC can be used to manipulate spin states and
its strength can be tuned by external gate voltage conveniently$^{3-5}$.

By introducing Rashba SOC in semiconductors nanostructures, several spin
filtering devices have been proposed without need for a magnetic element or
an external magnetic field, such as T-shape electron
waveguide$^{6-10}$, quantum wires$^{11-14}$, two-dimensional electron
gas (2DEG)$^{15}$, and quantum rings$^{16}$. Recently, Zhai et al. have proposed a spin
current diode based on a hornlike electron waveguide with Rashba SOC and
found that a quite different magnitude of spin-polarized current can be achieved when the
transport direction is reversed$^{17}$. The physical mechanism
of the proposed device arises from spin-flipped
transitions caused by the spin-orbit interaction. However, only transversal
spin conductance could be nonvanishing in the considered system, duo to its mirror
symmetry along the transverse direction is broken but respect to the
longitudinal direction is remained. The
spin-polarized transport properties of a Rashba step-like quantum wire
is also investigated by Xiao and Chen$^{18}$. It is shown that a very large
spin conductance can be obtained when the forward bias is applied to
the structure, while it is vanished or suppressed
when the transport direction is reversed . This effect is owing to
the different local density of electron states in the quantum wire
when the transport direction is reversed. However, the two
types mirror symmetries (longitudinal and transverse direction symmetries) of
the step-like quantum wire are all destroyed, and the $C_2$ rotation symmetry$^{19}$ will also be
invalid. Thus it is not known whether a large spin-polarized current
could be generated when the two types mirror symmetries of the investigated
system are all destroyed, but the $C_2$ rotation symmetry is remained. Further, the
influence of the disorder for a real application, remains unclear.

Inspired by the two works above, in this paper, we study the spin-dependent
electron transport for an asymmetrical quantum wire, in which transversal and
longitudinal symmetries are all broken, however, the $C_2$ rotation symmetry is reserved.
It is found that the magnitude of the spin-polarized current in the backward biased case is equal to
it in the forward biased case but their signs are contrary. This
results indicate that the spin polarization direction
can be switched by changing the direction of the external current. Furthermore, this
spin-polarized current can survive even in the presence of strong disorder. Therefore, a
spin polarization direction switch device can be devised by using this system.

The system investigated in present work is schematically depicted in Fig.
1, where a 2DEG in the $(x,y)$ plane is restricted to a an asymmetrical quantum
wire by a transverse confining potential $V(x,y)$. The SOC is assumed to
arise dominantly from the Rashba mechanism since the 2DEG is
confined in a asymmetric quantum well. The asymmetrical QW consists of two
regions, the left and right regions have the same length $L_{1}$ and
a uniform width $W_{1}$. The two connecting leads are normal-conductor
electrodes without SOC since we are only interested in spin-unpolarized
injection. We choose the coordinate system such that the $x$ axis, with $l$
lattice sites, is in the longitudinal direction, while
the $y$ axis, with $m$ lattice sites, is in the transverse
direction. To describe the electronic properties of the effective
discretized system of square lattice, one can define the tight-binding
Hamiltonian including the Rashba SOC on a square lattice as follows$^{20}$
\begin{eqnarray}
H=\sum\limits_{lm\sigma}\varepsilon_{lm\sigma}C_{lm\sigma}^{\dag}C_{lm\sigma}-t\sum\limits_{lm\sigma}\{C_{l+1m\sigma}^{\dag}C_{lm\sigma}
+C_{lm+1\sigma}^{\dag}C_{lm\sigma}+H.c\}\nonumber\\+t_{so}\sum\limits_{lm\sigma\sigma'}\{C_{l+1m\sigma'}^{\dag}(i\sigma_{y})_{\sigma\sigma'}C_{lm\sigma}
-C_{lm+1\sigma'}^{\dag}(i\sigma_{x})_{\sigma\sigma'}C_{lm\sigma}+H.c\}\nonumber\\+\sum\limits_{lm\sigma}V_{lm}C_{lm\sigma}^{\dag}C_{lm\sigma},
\end{eqnarray}
in which $C_{lm\sigma}^{\dag}(C_{lm\sigma})$ is the creation
(annihilation) operator of electron at site $(lm)$ with spin
$\sigma$($\sigma=\uparrow,\downarrow$), and $\sigma_{x}$ and $\sigma_{y}$ are
Pauli matrix. The on-site energy $\varepsilon_{lm\sigma}=4t$ with the hopping
energy $t=\hbar^{2}/2m^{\ast}a^{2}$,  here $m^{\ast}=0.067m_0$ is effective mass of electron
and $a$ is lattice constant. The Rashba SOC is
$t_{so}=\alpha/2a$ with the Rashba constant $\alpha$. $V_{lm}$ is
the additional confining potential. The Anderson disorder can be introduced
by the fluctuation of the on-site energies, which distributes randomly
within the range width $w$ [$\varepsilon_{lm\sigma}= \varepsilon_{lm\sigma}+w_{lm}$ with
$-w/2<w_{lm}<w/2$].

The spin-dependent conductance from arbitrary lead $p$ to lead $q$ is given by$^{18}$
\begin{eqnarray}
G^{\sigma'\sigma}=e^2/hTr[\Gamma_{N}^{\sigma}G_{r}^{\sigma\sigma'}\Gamma_{W}^{\sigma'}G_{a}^{\sigma'\sigma}],
\end{eqnarray}
where $\Gamma_{N(W)}=i[\sum_{N(W)}^{r}-\sum_{N(W)}^{a}]$ with the
self-energy from the narrow (wide) lead
$\sum_{N(W)}^{r}=(\sum_{N(W)}^{a})^{\ast}$, the trace is over the
spatial degrees of freedom, and
$G_{r}^{\sigma\sigma'}(G_{a}^{\sigma'\sigma})$ is the retarded
(advanced) Green function of the whole system that can be computed
by the spin-resolved recursive Green function method$^{21}$.

The local density of electron states (LDOS) is described as$^{22}$
\begin{eqnarray}
\rho(\vec{r},E)=-\frac{1}{2\pi}A(\vec{r},E)=-\frac{1}{\pi}Im[G_r(\vec{r},E)],
\end{eqnarray}
where $A\equiv i[G_r-G_a]$ is the spectral function, and $E$ is the
electron energy. In the calculations, the asymmetrical QW is taken to be that in a 2DEG
of high-mobility $GaAs/Al_{x}Ga_{1-x}As$ with a typical electron
density $n\sim 2.5\times 10^{11}~/cm^{2}$, all the energy is normalized
by the hoping energy $t(t=1)$. The structural parameters of the asymmetrical QW
are fixed at $L_1=10~a$, $W_1=10~a$ and $L_2=6~a$, with $a$ lattice
spacing of the tight-binding model. And the $z$ axis is chosen as
the spin-quantized axis so that $|\uparrow>=(1,0)^{T}$ represents
the spin-up state and $|\downarrow>=(0,1)^{T}$ denotes the spin-down
state. The boundary of the wire is determined by the hard-well
confining potential. The total charge conductance and the spin
polarization of $z$-component are defined as
$G^{e}=G^{\uparrow\uparrow}+G^{\downarrow\uparrow}+G^{\downarrow\downarrow}+G^{\uparrow\downarrow}$
and
$P_{z}=((G^{\uparrow\uparrow}+G^{\uparrow\downarrow})-(G^{\downarrow\downarrow}+G^{\downarrow\uparrow}))/G^{e}$,
respectively.

Figure 2(a) shows the total charge conductance as function of the
electron energy when the spin-unpolarized electrons are
injected into the considered system from the Left (L) lead to the Right (R)
lead (the forward biased case). When $E<0.10$, the charge conductance
is zero since all the subbands of the quantum wire are evanescent modes.
The step-like structures are formed in the charge conductance
with each step height being $2$ in the straight wire$^{19}$. But, when
the electron energy $E>0.10$, in the asymmetrical QW unlike the
straight wire, the charge conductance fluctuates in each step and
the average heights of those steps are reduced, due to the scattering
at interfaces between the left and right quantum wire of the investigated
system. Furthermore, oscillations also emerge in the total charge
conductance, which may result from the multi-reflection at the interfaces between
the left and right quantum wire of the asymmetrical QW. The oscillation
periodicity is related to the wave vectors of the propagating modes so that the oscillations
become apparent just above the thresholds of the subbands where the
wave vectors turn out to be smaller$^{23}$. The SOC-induced Fano
resonance dips also can be found in the charge conductance. The corresponding
spin polarization of Fig. 2(a) as show in Fig. 2(b), when $0.10<E<0.39$, only
the lowest one pair of subbands of the quantum wire are propagating
modes, so there is no spin polarized current$^{24}$. But
when $E>0.39$, both the inputting lead and outgoing lead support
two or more pairs of propagating modes and the subband intermixing
induced by the Rashba SOC arises, resulting in the nonzero spin
polarized current. It is worth to note that a valley-like
structure [see the red rectangles in Fig. 2(a)] appears in the
charge conductance when the emitting energy just near
the threshold of the third pair of propagating modes in the right quantum
wire, i.e., $E=0.83$ and $E=1.36$. This effect may be attributed
to the bound state in the quantum wire couples to the conductance
one, giving rise to a structure-induced Fano resonance$^{21}$. Amazingly, at
one of those Fano resonance (such as $E=0.83$), a large spin polarization $|P_{z}|=0.30$ can
be achieved in the outing lead. This effect may result from the
two types of mirror symmetries (longitudinal and transverse direction symmetries) are all
broken in the asymmetrical QW. Moreover, with the presence
of Rashba SOC in the investigated system, the transparency of an initial spin-up
electron is no longer equal to that of an initial spin-down electron. As a
result, the spin polarization of the $z$-component is nonzero.

Figure 2(c) plots the total charge conductance as a function of the
electron energy when electrons are injected into the considered structure from
the lead R to L (the backward biased case). The total
charge conductance is the same as in Fig. 2(a) due to the
space-inversion symmetry of the asymmetrical QW. The corresponding spin polarization
of Fig. 2(c) is plotted in Fig. 2(d). Interestingly, the magnitude of the
spin polarization is equal to it in Fig. 2(b) but their signs are reversed. This
may due to these two types of mirror symmetry are both destroyed, but the
$C_2$ rotation symmetry is still reserved. Together with the
time-reversal symmetry, we obtain the relation $G^{\uparrow\uparrow}_{LR}=G^{\downarrow\downarrow}_{RL}$,
$G^{\uparrow\downarrow}_{LR}=G^{\downarrow\uparrow}_{RL}$, $G^{\downarrow\downarrow}_{LR}=G^{\uparrow\uparrow}_{RL}$ and
$G^{\downarrow\uparrow}_{LR}=G^{\uparrow\downarrow}_{RL}$ [see Figs. 3(a) and 3(b)]. This means that
for the forward and backward biased case, the magnitude of the spin polarization is equal,
but their signs of the spin polarization are contrary. The characteristics
in the spin polarization between the forward and backward transport
directions can be utilized to devise a spin polarization direction switch
device, in which the spin polarization direction can be switched by changing
the direction of the external current, that is it can be rectified by all-electrical method.

In order to clarify this effect, the spin-dependent conductance for the
forward and backward biased case as function of the electron
energy is illustrated in Figs. 3(a) and 3(b), respectively. The strength
of Rashba SOC $t_{so}=0.177$. As show in Fig. 3(a), the spin-polarized
components $G^{\uparrow\uparrow}_{LR}(G^{\downarrow\downarrow}_{LR})$ and
$G^{\uparrow\downarrow}_{LR}(G^{\downarrow\uparrow}_{LR})$ exhibit a series
of resonant structures because the Rashba SOC induces the spin splitting and
results in a subband intermixing$^{25}$. At the energy near the bottom
of the second to fourth subband ($E=0.39$, $0.83$, and $1.36$, respectively), the spin conductances exhibit a sharp peak
for $G^{\uparrow\uparrow}_{LR}(G^{\downarrow\downarrow}_{LR})$ and a sharp dip for
$G^{\uparrow\downarrow}_{LR}(G^{\downarrow\uparrow}_{LR})$ in the forward
biased case. This phenomenon is related to the details of the spin-dependent
scattering mechanism of electron transport through the whole system
configuration$^{13}$. Due to the longitudinal and transversal symmetries are all broken in the asymmetrical QW, so
the relations $G^{\uparrow\uparrow}_{LR(RL)}=G^{\downarrow\downarrow}_{LR(RL)}$
and $G^{\uparrow\downarrow}_{LR(RL)}=G^{\downarrow\uparrow}_{LR(RL)}$ cannot
be guaranteed, leading to the nonzero spin polarization[as show in Figs. 2(b)
and 2(d)] for the forward and backward biased case, respectively. Furthermore, the
Hamiltonians of the asymmetrical QW are also invariant under $C_2$ rotation. Therefore, from
the $C_2$-rotation and time-reversal symmetries, we can find that the transmission
probability of the spin-up (-down) electron from the lead L to R always equals that
of the spin-down (-up) electron from the lead R to L. As a consequence, the magnitude
of the spin polarization in the backward biased case is equal to it in the forward
biased case and their signs are reversed as show in Figs. 2(b) and 2(d).

The LDOS of the asymmetrical QW for the forward and backward biases
is shown in Figs. 5(a) and 5(b), respectively. The electron
energy is taken to be $E =0.83$, and the strength of Rashba
SOC $t_{so}=0.177$. As shown in Fig. 4(a), for the studied system
in the forward biased case, two regular stripe appears in the left
region of the asymmetrical QW that represents two pairs
of propagating modes, whereas an obvious bound state is found to
exist in the top of the right region of the asymmetrical QW. In this
case, electrons are equivalent to transmit from a potential barrier
region (the left region of the asymmetrical QW) to a potential well
region (the right region of the asymmetrical QW) due to the transversal
confining potential$^{18}$. Thus electrons have a certain probability to
stay in the right region of the asymmetrical QW. This is due to in
the asymmetrical QW, different transverse modes can be
mixed by corner scattering as well as by interface scattering at
the interfaces between the left and right quantum wire of the investigated system. As
a consequence, higher-index propagating modes are preferred to be
populated inside the right region of the quantum wire$^{7}$. Moreover,
this bound state interacts with the Rashba SOC-induced effective
magnetic, resulting in the transmission coefficient of the spin-up
electron is injected from the inputting lead can be very different from
that of the spin-down electron. Therefore, a large spin polarization can be achieved in the
outgoing lead. However, in the backward biased case, as shown in
Fig. 4(b), two regular stripe appears in the right region of the
asymmetrical QW and an obvious bound state is found to exist in the
bottom of the right region of the asymmetrical QW. This may due to the
$C_2$-rotation and time-reversal symmetries in the asymmetrical QW are all remained.

The above proposed spin polarization switch device is based on a
perfectly clean system, where the influence of impurity scattering
is not taken into account. But usually, in a realistic system
contains a lot of impurities, and they are distributed
randomly. Thus the effect of disorder should be considered in
practical application. Now we show the feasibility of this device
for a real application by analyzing the robustness of the spin
conductance against the Anderson disorder. The total charge
conductance and corresponding spin polarization as a function of the electron
energy for different disorders $w$ are illustrated in Fig. 5. The
Rashba SOC strength $t_{so}=0.177$. The step-like charge
conductance is destroyed with the increase of the disorder strength
as show in Figs. 5(a) and 5(c), respectively. However, as shown in the lower panel
in Fig. 5(b), the magnitude of the spin polarization around the
thresholds of the propagating modes in the asymmetrical QW is still
large, which indicates that the spin polarization can still survive
even in the presence of strong disorder when electrons are injected from
the forward direction. At the same time, the spin polarization is
still very large for a strong disorder when electrons are injected from
the backward direction, as shown in the lower panel in Fig. 5(d). The
underlying physics could also be attributed to the different LDOS
for the diverse bias directions even in the presence of disorder, as
shown in Fig. 6. The disorder strength $w=0.4$ and the other parameters
are the same as that in Fig. 4. By comparing with
Figs. 4(a) and (b), the LDOS is redistributed in the quantum wire due to
the disorder-induced scattering. However, the salient features
remain the same as that in Figs. 4(a) and (b).

In summary, a scheme of spin polarization direction switch is
proposed by investigating the spin-dependent electron transport
of an asymmetrical QW with the modulation of the Rashba
SOC. It is shown that the magnitude of the
spin-polarized current in the backward biased case is equal to
it in the forward biased case but their signs are contrary. The underlying physics is revealed to
originate from the symmetries in the longitudinal and transverse directions
are all broken but $C_2$-rotation and time-reversal symmetries are
still reserved. Further studies show that the spin polarization
is robust against disorder, This results may provide an efficient
method to generate an artificially controllable spin polarization
direction in mesoscopic Rashba systems without applying an
external magnetic field and without attaching ferromagnetic contacts.

This work was supported by the National Natural
Science Foundation of China(Grant No.10774112), and the
National Natural Science Foundation of China under(Grant No.11147156).

\newpage

\begin{figure}
\caption{\label{fig:wide}Schematic diagram of the asymmetrical
quantum wire with Rashba SOC. the left and right quantum wires
have the same length $L_1$ and a uniform width $W_1$.}
\end{figure}

\begin{figure}
\caption{\label{fig:wide} (Color online) The calculated total
charge conductance and the corresponding spin polarization as
a function of the electron energy. (a) The forward biased
case. (b) The backward biased case. The Rashba SOC strength $t_{so}=0.177$.}
\end{figure}

\begin{figure}
\caption{\label{fig:wide} (Color online) The calculated
spin-dependent conductance as function of the electron
energy when the spin-unpolarized electron is injected. (a) The
forward biased case. (b) The backward biased case. The
Rashba SOC strength $t_{so}=0.177$.}
\end{figure}

\begin{figure}
\caption{\label{fig:wide} (Color online) The calculated
LDOS of the quantum wire in the forward biased case (a) and
in the backward biased case (b). The strength of Rashba
SOC $t_{so}$ is fixed at $0.177$. The electron
energy $E=0.83$. The arrow denotes the transport direction of electrons.}
\end{figure}

\begin{figure}
\caption{\label{fig:wide} (Color online) The calculated
total charge conductance and the corresponding spin polarization
as a function of the electron energy for different disorder
strengths. (a) The forward biased case. (b) The backward
biased case. The Rashba SOC strength $t_{so}=0.177$.}
\end{figure}

\begin{figure}
\caption{\label{fig:wide} (Color online) The calculated LDOS
of the disordered asymmetrical quantum wire in the forward
biased case (a) and in the backward biased case (b). The
strength of disorder $w=0.4$, and other parameters are the
same as that in Fig. 4. The arrow denotes the transport direction of electrons.}
\end{figure}

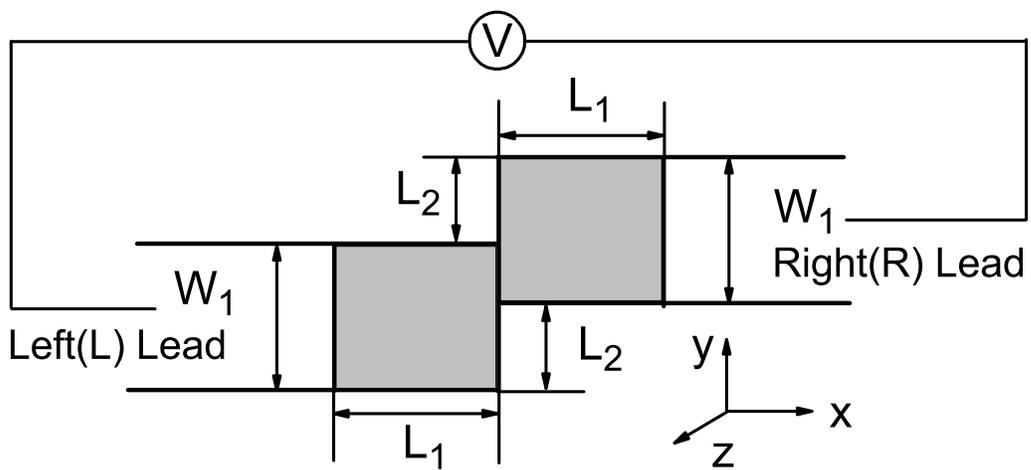
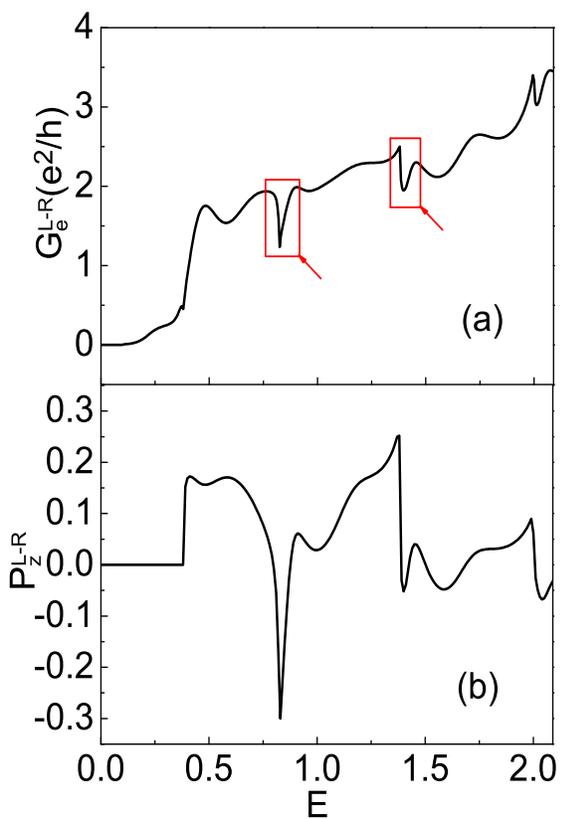
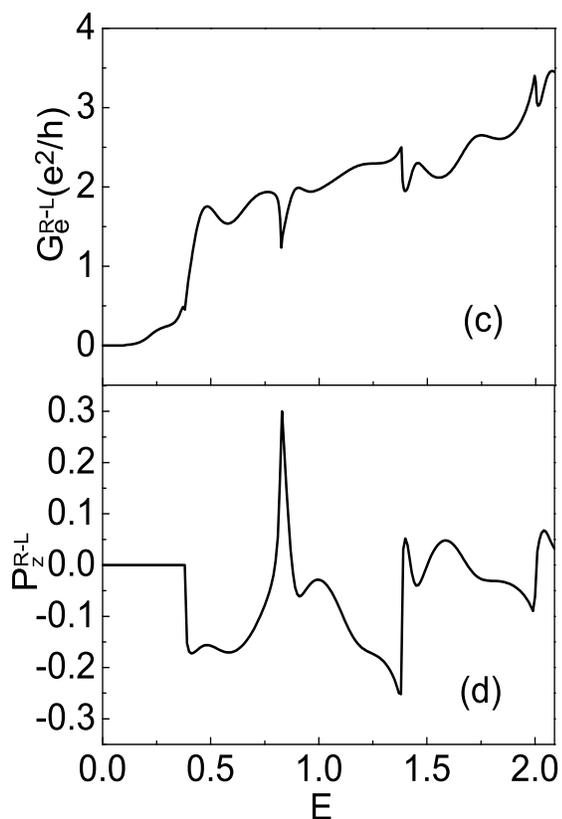
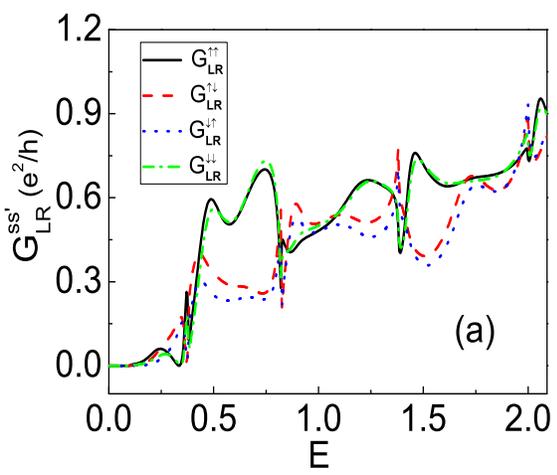
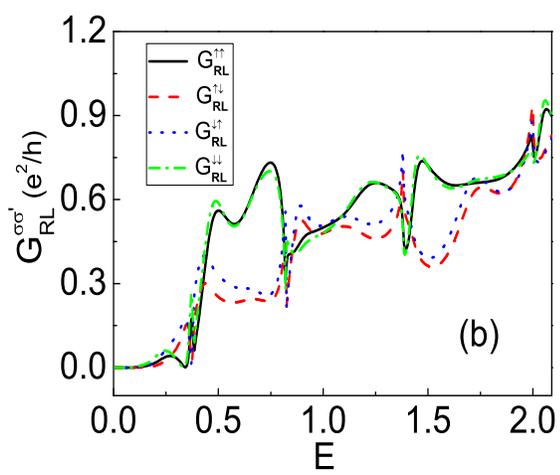
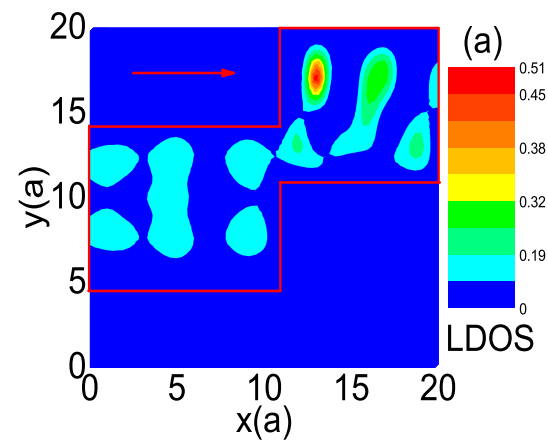
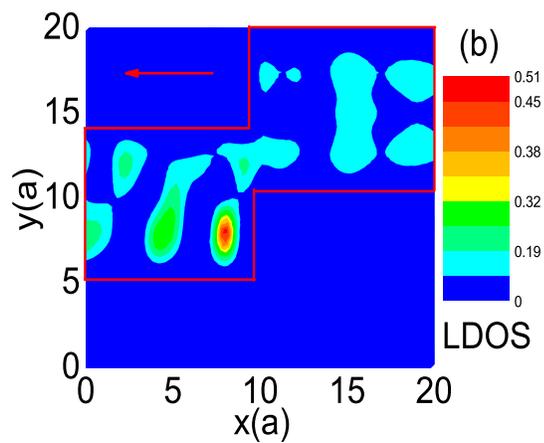
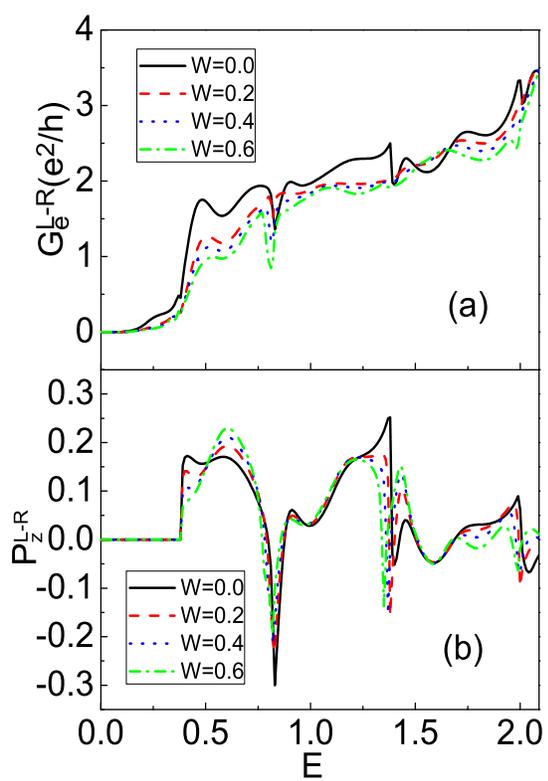
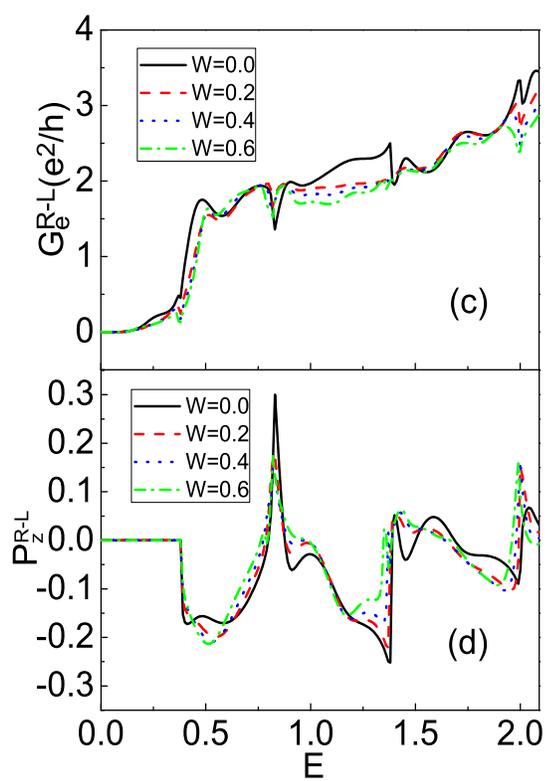
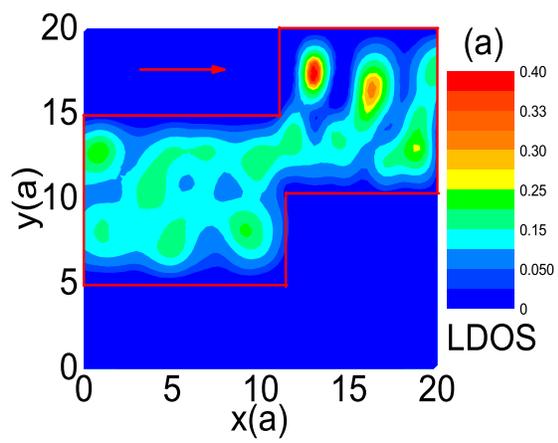
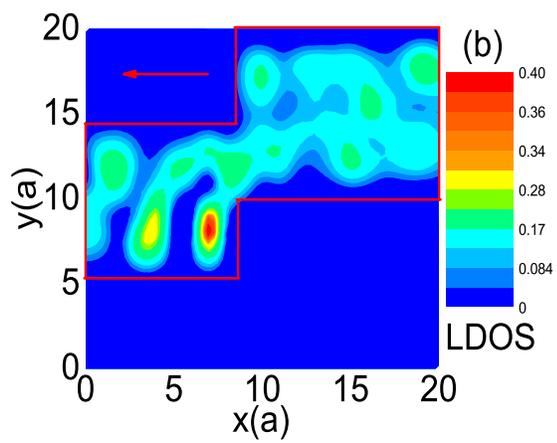
\end{document}